\documentclass[11pt,aps,preprint]{revtex4}
\usepackage{graphicx}
\usepackage{amssymb}
\usepackage{epstopdf}
\begin{document}
\title{Non-observability of Spectroscopic Factors}
\author{B.K. Jennings}
\email{jennings@triumf.ca}
\affiliation{TRIUMF, 4004 Wesbrook Mall, Vancouver, BC, V6T 2A3}\vspace{2cm}

\date{\today}                                           
\begin{abstract}
The spectroscopic factor has long played a central role in nuclear reaction theory. However, it is not an observable. Consequently  it is of minimal use as a meeting point between theory and experiment. In this paper the nature of the problem is explored. At the many-body level, unitary transformations are constructed that vary the spectroscopic factors over the full range of allowed values.  At the phenomenological level, field redefinitions play a similar role and the spectroscopic factor extracted from experiment depend more on the assumed energy dependence of the potentials than on the measured cross-sections. The consistency conditions, gauge invariance and Wegmann's theorem play a large role in these considerations.
\end{abstract}

\maketitle

\section{Introduction}

The spectroscopic factor, $S$, has a long history of use in nuclear and other many-body problems. The hope has been that it would provide a meeting ground between theory and experiment where structure effects could be included in reaction calculations in a simple manner. In practice, there have been worries, in particular, with the inherent model dependence of the spectroscopic factors extracted from experiment. A more recent concern is that spectroscopic factors calculated from soft potentials tend to be closer to one ($S\approx 0.9$)\cite{Jensen} than those from the traditional potentials with strong short-range repulsion ($S\approx0.65$)\cite{DB}. 

The first attack on the validity of using spectroscopic factors dates from 1979 \cite{amado,friar} when it was shown that the d-state probability of the deuteron, a special case of the spectroscopic factor\cite{FS}, was not an observable. Its variation with a momentum cutoff is shown in Fig.~57 of ref.\cite{BFS} and, as can be seen from that figure, it goes to zero as the cutoff gets stronger. The 1979 papers ended attempts to experimentally determine the d-state probability. The spectroscopic factor is related to the off-shell behaviour of the NN potential. The off-shell behaviour of the NN was shown to be unobservable, at least in NN bremsstrahlung, in 1998\cite{fearing}. Following this paper, attempts to use bremsstrahlung to measure off-shell properties largely ceased. A more direct attack on the measurability of spectroscopic factors occurred in 2002\cite{FH} when it was shown that spectroscopic factors, in general, are not observables. Inexplicably, this has had little effect on the efforts to measure or calculate spectroscopic factors. More recently, it was shown\cite{muk} that spectroscopic factors are not invariant under finite-range unitary transformations. This explains the difference observed in the calculated spectroscopic factor with hard and soft potentials. In all the cases cited above, the ambiguity arises from a shuffling of contributions from the wave functions to the transition operators or equivalently from the nuclear structure to the reaction mechanism. 

In the present paper, it will be shown that spectroscopic factors are only meaningful within the context of the model used to derive or calculate them. In theoretical calculations (Sect.~\ref{sec:mbp}), they depend on the model and the representation used. This is more general than the cut-off dependence in an effective field theory. The cutoff is  one dimension in the $(N-1)^2$ dimensional space of non-trivial unitary transformations in an $N$ dimensional Hilbert space.  An explicit unitary transformation is constructed that makes all the spectroscopic factors either zero or one. In Sec.~\ref{sec:obp}, a derivation of an effective one-body potential and transition operator are given using the particle-hole formalism. This sets the stage for a discussion of phenomenological approaches (Sec.~\ref{sec:ph}) for extracting the spectroscopic factor, as these are usually based on one-body equations. The phenomenologically extracted spectroscopic factors depend on the potentials and transition operators that are used to extract them and, as a matter of principle, these are not uniquely defined. Reasonable variations in the potentials cause significant changes to the extracted spectroscopic factors without affecting observables. In showing this it is necessary to go beyond a pure DWBA calculation since consistency with gauge invariance and Wegmann's theorem\cite{wegmann} (see eq.~\ref{eq:weg} below) implies $S=1$ for a pure DWBA calculation. In Sec.~\ref{sec:po}, additional insights from considering a particle-only formalism are presented.  Finally the conclusions are given in Sec.~\ref{sec:con}. In the present paper, the effect of nuclear recoil are neglected (see ref.~\cite{EJS} for details on the recoil correction).

\section{Spectroscopic factors and the many-body problem}
\label{sec:mbp}

The basic problem with the spectroscopic factor goes back to its definition. It is not defined in terms of a conserved current or even a matrix element corresponding to an observable. Rather it is defined in terms of the form of the matrix element. In second-quantized notation:
\begin{eqnarray}
S&=& \int d^3r | \phi(r)|^2, 
\end{eqnarray}
where
\begin{eqnarray}
\phi(r) &=& \langle \Phi_{A-1} | a(r) | \Phi_A \rangle.
\end{eqnarray}
The crucial point in the definition is that it is the matrix element of the quasi-particle destruction operator. Under a unitary transformation, $a(r) \Rightarrow U^\dag[a(r'),a^\dag(r')] a(r) U[a(r'),a^\dag(r')]$, the form is changed so it is no longer just a single destruction operator but in general a complicated function of creation and distruction operators. Hence, it no longer part of the definition of a spectroscopic factor. It is the spectroscopic factor normalized to one that can be identified with the phenomenological single-particle wave function (see Sec.~\ref{sec:obp}). 

The argument is clearer in the first-quantized representation:
\begin{eqnarray}
\phi(r) &=&\sqrt{A} \int \prod_{i=1}^{A}dr_i \Phi^\dag_{A-1}(r_1,\ldots,r_{A-1} ) \delta(r-r_A)\Phi_A(r_1,\ldots,r_{A})
\end{eqnarray}
The crucial point in defining the spectroscopic amplitude, $\phi(r)$, is  the form with the $(A-1)$-body wave function times a delta function. If we insert $1 = U^\dag_AU_A$ to the left of $\Phi_A$ and have $U_A^\dag$ act to the left, the form changes even for trivial 
transformations. A one body-transformation $U_A=\prod_{i=1}^A U_i$ will change the delta function to $U_i(r-r_A)$, meaning that in the new representation we no longer have the spectroscopic amplitude (although in this simple case the spectroscopic factor does not change). Many-body transformations, such as that in the  Unitary Correlated Operator Method (UCOM)\cite{feldmeier}, will destroy the product form and change the spectroscopic factor. 

Now consider what happens if we keep the form fixed:
\begin{eqnarray}
\phi_{UT}(r) &=&\sqrt{A} \int \prod_{i=1}^{A}dr_i \Phi^\dag_{A-1}(r_1,\ldots,r_{A-1} ) \delta(r-r_A)U^\dag_{A-1} U_A\Phi_A(r_1,\ldots,r_{A}).
\end{eqnarray}
Note that we have a $(A-1)$-body transformation on the left and an $A$-body transformation on the right. This is normally what is meant when we talk about how the spectroscopic factor transforms under unitary transformations (see for example ref.~\cite{muk}).  Even for relatively simple forms of the unitary transformation the norm of $\phi_{UT}(r)$ changes. For example, $U = \exp[\sum_{i<j}c \cdot \nabla_{ij}]$, where the sum runs over all particles, transforms the spectroscopic amplitude to:
\begin{eqnarray}
\phi_{UT}(r) &=&\sqrt{A} \int \prod_{i=1}^{A}dr_i \Phi^\dag_{A-1}(r_1-c,\ldots,r_{A-1}-c ) \delta(r-(A-1)c-r_A)\Phi_A(r_1,\ldots,r_{A})
\end{eqnarray}
If the many-body wave functions are localized in space the translation will reduce the overlap as the unitary transformation moves the $(A-1)$-body cluster on the left away from the $A$-body system on the right. In this case, as $c$ becomes large the overlap and therefore the spectroscopic factor, will go to zero. For example, consider a one dimensional four-body system where $\Phi(r_1,\ldots,r_4)= \prod_{i=1}^4 (2/\pi)^{1/4}\exp[ - r_i^2/r_0^2] $ and a three-body state that is the same except that the product only goes from one to three. The spectroscopic amplitude becomes $ (2/\pi)^{1/4}\exp[-(r+c)^2/r_0^2] \exp[- 3 c^2/r_0^2]$ and the spectroscopic factor goes as $\exp[- 6 c^2/r_0^2]$. Both are explicitly dependant on the parameter $c$ from the unitary transformation. Thus we can see in this simple case how a unitary transformation can change the spectroscopic factor. This is special case of the more general considerations of ref.~\cite{muk}.

Having established that unitary transformations can change the spectroscopic factors, we might ask if the size of the change is constrained.  A unitary transformation can always be written as $U = \sum_i |\Phi_i\rangle\langle\Psi_i |$ where the $|\Phi_i\rangle$ and $|\Psi_i\rangle$ are complete orthonormal sets that span the space of interest: $|\Phi_i\rangle = U |\Psi_i\rangle$.  Let us take the space of interest to be the $(A-1)$, $A$ and $(A+1)$ many-body states, the $|\Psi_i\rangle$ to be the eigenstates of the many-body Hamiltonian in the original representation and the $|\Phi_i\rangle$ to be Slater determinants built with a common set of orbitals, for example the orbitals generated by the Hartree-Fock potential in the $A$-body system. The new states are product states so the spectroscopic factors will be either zero or one depending on which orbitals are occupied in the Slater determinants. Thus, by construction we have shown that unitary transformations exist where the spectroscopic factors for the transformed states take the extremal values, zero or one. With a bit of ingenuity the transformation can be modified to give intermediate values. Thus the unitary transformations can vary the spectroscopic factors at will subject only to general principles like the spectroscopic factors must be between zero and one (neglecting centre-of-mass corrections). 

Strictly speaking the simple analysis of the last paragraph is only valid if all the states are bound states. For scattering states there are additional constraints\cite{ekstein,polyzou} on the asymptotic properties of the unitary transformation. The scattering states are normalized asymptotically so they do not have a spectroscopic factor associated with them. Thus it is only useful to apply the above transformation to bound states. The short range parts of the scattering states can then be adjusted so they are orthogonal to the new bound states by using a variation of the Lee-Suzuki method\cite{LS}cite. 

It might be argued that the above unitary transformation is rather arbitrary. However, all observables are unchanged under even the most arbitrary unitary transformation. Some other criteria than agreement with experiment would have to be invoked to rule out such transformations. The above transformation, however, might even be useful since, if we take the orbitals to be those generated by a Hartree-Fock calculation in the $A$-body system, all it does is transfer the effects of correlations from the wave functions to the operators.

The question now arises: How does the representation dependence of the spectroscopic factor manifest itself in phenomenological determinations of the spectroscopic factor. This will be dealt with in the following sections. First, we establish how the phenomenological models arise from the microscopic models and explore the ambiguities in the phenomenological models. It will be shown that the spectroscopic factor extracted from experiment can be varied arbitrary by the effective one-body equivalent of the unitary transformation: the field redefinition. 

\section{The Effective One-Body Problem: The Particle-Hole Approach}
\label{sec:obp}

Attempts to extract spectroscopic factors typically depend on the DWBA approximation. This, in turn, relies on reducing the many-body problem to an effective one-body problem. So it is here we start. Consider, for example, a proton scattering off an oxygen nucleus. The phenomenological potential that is used to describe this process, for example a Woods-Saxon, usually has bound states that correspond to states of both Fluorine, eg.\ the $5/2^+$ state, and Nitrogen, eg.\ the $3/2^-$ and $1/2^-$ states. Thus a derivation of the effective one-body potential will have to include both particle and hole states.  The particle-hole approach, based on the self-energy operator, is discussed extensively in ref.~\cite{MS} where it is argued convincingly that it is the self-energy operator that the phenomenological potentials approximate. In this paper, we follow the techniques of ref.~\cite{EJ},  rather than of ref.~\cite{MS}, to derive expressions for the self-energy operator and effective transition operators. The self-energy operator is the same in either approach.

Since we need both particle and hole states, we start with the particle-hole projection operator: 
\begin{eqnarray}
P_{PH}&=&\int d^3r [a(r) +a^\dag(r)]|\Psi_A\rangle\langle \Psi_A| [ a(r)+a^\dag(r)]\label{eq:p}.
\end{eqnarray}
In the particle-hole space, the spectroscopic amplidute is defined as:
$\phi(r) = \langle \Psi_A |  [a(r) +a^\dag(r)]|\Psi_{(A-1),(A+1)} \rangle$
where the state on the right can have either $(A-1)$ or $(A+1)$ particles.  In either the particle or hole space this definition of the spectroscopic amplitude reduces to the usual one. We use the particle-hole projection operator, $P_{PH}$,  to define a model space and use the standard model-space techniques. These techniques are more commonly used with a projection onto a model space defined by the energy as with the shell model, but the techniques are general. We could use the Lee-Suzuki\cite{LS} approach. This uses a unitary transformation and all the states in the model space defined by $P_{PH}$ have spectroscopic factor $1$ while those in the $Q_{PH}=1-P_{PH}$ space would have spectroscopic factor $0$. This indicates, yet again, that the spectroscopic factor is not physical but rather an artifact of the methods used. The effective Hamiltonian is energy independent in this approach, consistent with Wegmann's theorem (see below). 

Now we turn to the Feshbach projection operator formalism\cite{Feshbach}. From ref.~\cite{EJ} we have:
\begin{eqnarray}
{\cal E} \phi(r) &=& \int dr'\langle \Psi_A |[a(r)+a^\dag(r)]
\left( \tilde H + \tilde H Q_{PH} \frac{1}{{\cal E} - Q_{PH}\tilde H
Q_{PH}}Q_{PH} \tilde H\right)\nonumber 
\\&& \hspace{1cm} 
\times [a(r')+ a^\dag(r')]| \Psi_A^n\rangle\phi(r') 
\\&\equiv& \int dr' {\cal H}_{\cal M} (r,r' ,{\cal E})\phi(r'),
\label{eq:massab}
\end{eqnarray}
where ${\cal E}=(E^{A+1}-E^A)$ 
holds for particle states, ${\cal E} = (E^A-E^{A-1})$
 holds for hole states,  $\tilde H
\equiv (H-E^A_0)(\hat A - A)$, $\hat A$ is the number operator, and  ${\cal H}_{\cal M} (r,r' ,{\cal E})$ is the particle-hole self-energy operator as in ref.~\cite{MS}. Note that it is energy dependent. If the energy is small compared to $Q_{PH}\tilde H Q_{PH} $ we can expand in powers of the energy. Thus large values of $Q_{PH}\tilde H Q_{PH}$ give constant and linear terms in the energy. As we will see in the next section, the field redefinition also introduces a term linear in the energy into the potential. The linear term here suggests that the linear term there might also be related to high energy components in many-body Hamiltonian. 

Wegmann's theorem\cite{wegmann} relates the spectroscopic factor to the energy dependence of the effective one-body Hamiltonian:
\begin{eqnarray}
S=\int dr\; |\phi(r)|^2= \left[ 1-\frac{d}{d\mathcal E} \int
drdr'\; {\hat\phi(r)}^* {\cal H}_{\cal M} (r,r' ,{\cal E}) 
 \hat\phi(r')\right]^{-1}.\label{eq:weg}
\end{eqnarray}
where the carat, $\hat{\phantom{a}}$, denotes spectroscopic amplidutes normalized to one not $S$. This equation was originally derived by comparing the residues of the poles in the Green function based on two different representations. In the present approach, this can be derived from $\langle \Psi_A|(P+Q)| \Psi_A\rangle$ and eq.~\ref{eq:p} and the relation\cite{Feshbach}:
\begin{eqnarray}
Q_{PH} |\Psi_A \rangle = \frac{1}{{\cal E} - Q_{PH}\tilde H
Q_{PH}}Q_{PH}  \tilde H P_{PH} | \Psi_A \rangle.
\label{eq:q}
\end{eqnarray}

To calculate a transition we also need the matrix element $\langle \Psi_A^I|\theta| \Psi_A^J\rangle$ where $\theta$ is the transition operator. For electric transitions $\theta = \int d^3r\exp[ip\cdot r] a^\dag(r)a(r)$. Here we have assumed a one-body operator but in general there will be many-body operators coming from exchange currents. Even if the operator in the many-body space is simple, the operator in the one-body space will be complicated, going beyond the free one-body operator. Taking $\langle \Psi_A^I|(P_{PH}+Q_{PH})\theta (P_{PH}+Q_{PH}) | \Psi_A^J\rangle$ we have four terms, one diagonal in $P_{PH}$, the two cross-terms in $P_{PH}$ and $Q_{PH}$ and the term diagonal in $Q_{PH}$. From the definition of $P_{PH}$, the first diagonal term can be seen to give two contributions: $\int d^3r \phi^{I*} (r) \theta(r) \phi^J(r)$ and recoil terms (ie. $\approx 1/A$). The off-diagonal terms also vanish in the absence of recoil effects. This leaves the last, which from eq.~\ref{eq:q} gives a contribution:
\begin{eqnarray}
 \langle \Psi_A | P_{PH} \tilde H Q_{PH} \frac{1}{{\cal E} - Q_{PH}\tilde H
Q_{PH}}  \theta  \frac{1}{{\cal E} - Q_{PH}\tilde H
Q_{PH}}Q_{PH}  \tilde H P_{PH} | \Psi_A \rangle \approx  \langle \Psi | \frac{ d {\cal H}_{\cal M}}{d \mathcal E} \theta | \Psi_A \rangle 
\label{eq:dvde}
\end{eqnarray}
that is proportional to the expectation value of $d {\cal H}_{\cal M} /d \mathcal E$ to within questions of the ordering of the operators. As we will see in the next section, minimal substitution at the effective one-body level also gives a contribution proportional to the energy derivative of the potential,  again with operator ordering ambiguities.

\section{Phenomenology}
\label{sec:ph}

For definiteness we consider radiative capture. Phenomenologically, we have two basic equations:
\begin{eqnarray}
&&\sigma \propto S\left| \int d^3r \hat\phi_f^*(r) T \phi_i(r)\right |^2 \label{eq:sigma}\\
&&-\hbar^2 \nabla\frac{1}{2 m(r) } \nabla\phi(r) +  V(E,r)  \phi(r) = E\phi.\label{eq:sch}
\end{eqnarray}
These follow from the considerations of the last section when we neglect recoil effects. The carat, $ \hat{\phantom{a}}$, signifies that the spectroscopic amplitude (i.e. the wave function, $\phi(r)$) is normalized to one not the spectroscopic factor. The scattering state, $\phi_i(r)$, is normalized asymptotically so there is no spectroscopic factor associated with it. We also have two consistency conditions:
\begin{eqnarray}
&&T = e^{iq\cdot r} \left(1 - \frac{dV(E,r)}{dE}\right)\label{eq:T}\\
&&S = \frac{1}{\int d^3r \hat\phi^*(r) \left(1 - \frac{dV(E,r)}{dE}\right) \hat\phi(r) }.\label{eq:S}
\end{eqnarray}
When the potential is energy dependent, taking the transition operator to be just the first term in eq.~\ref{eq:T} violates gauge invariance.  The second term in eq.~\ref{eq:T} follows using minimal substitution to guarantee gauge invariance. This form is similar to what was suggested in the last subsection where the term, eq.~\ref{eq:dvde}, had the form of a derivative. While minimal substitution is a fairly crude tool for getting the transition operator, at the purely phenomenological level it is the only one available. The second consistency condition is Wegmann's theorem discussed previously.

From the second consistency condition, it follows that the potential must be energy dependent if the spectroscopic factor differs from one. It then follows that the DWBA, which in its pure form keeps only the first term in eq.\ref{eq:T}, is inconsistent with gauge invariance. Or turning the argument around: The DWBA implies that the potential is energy independent and hence $S=1$. Thus in a DWBA calculation, $(S-1)$ is a measure of the inconsistency in the calculation; that is, a measure of the violation of Wegmann's theorem.

Let us now consider a field redefinition:
\begin{equation}
\phi(r) = f(r) \varphi(r).
\end{equation}
To keep the phase shifts invariant we take $f(r)$ to go to one for large $r$ and to guarantee an inverse $f(r)$ must have no zeros. Using the new fields, the spectroscopic factor $\mathcal S = \int d^3r |\varphi(r)|^2$ is numerically different from the original $S$.  For example, if we take $f(r)$ to be constant for radii where the bound state spectroscopic amplitudes are finite and going to one at infinity, then we have $S = \mathcal S |f(0)|^2$.
The expression for the cross-section now becomes:
\begin{eqnarray}
\sigma  &\propto &\left| \int d^3r \varphi_f^*(r) f^*(r)T  f(r)\varphi_i(r)\right|^2 = \left| \int d^3r \varphi_f^*( r)\mathcal T \varphi_i(r)\right|^2
\label{eq:ss}
\end{eqnarray}
where the second equality defines $\mathcal T$ and the spectroscopic factor is implicit in the spectroscopic amplitude (no carat, $ \hat{\phantom{a}}$, on the amplitude). The Schodinger equation, the equivalent of eq.~\ref{eq:sch}, is now:
\begin{eqnarray}
&&-\hbar^2f^*(r) \nabla\frac{1}{2 m(r) } \nabla f(r) \varphi(r) +[ f^*(r) V(E,r)f(r) + E(1 - |f(r)|^2 ) ] \varphi(r) = E\varphi 
\end{eqnarray}
The prefactor of $f^*(r)$ is necessary to make the Hamiltonian hermitian. When $f(r)$ is real, its contribution to the kinetic energy can be absorbed into a redefined effective mass, $\mathcal M(r)$, and energy independent terms that depend on the derivatives of $f(r)$. When $F(r) = \sqrt{m(r)}/m $, $\mathcal M(r)$ is just the constant $m$. The new potential is:
\begin{eqnarray}
\mathcal V(E,r)= f^*(r) V(E,r)f(r) + E(1 - |f(r)|^2 ) +\ldots 
\end{eqnarray}
where the terms not written out explicitly involve derivatives of $f(r)$ and are energy independent. Notice the term linear in the energy: $E(1 - |f(r)|^2 )$. By Wegmann's theorem a change in energy dependence is required if the spectroscopic factor changes and this term provides the required energy dependence even when the original potential is energy independent. Similarly when used with minimal substitution it generates the transition operator $\mathcal T$ required by eq.~\ref{eq:ss}.

Consider how this works in a simple case. Take $m(r) = m$ and the potential $V(r)$ to be energy independent. By Wegmann's theorem the spectroscopic factor is one and the transition operator is just $\exp[iq\cdot r]$and we expect the DWBA to be valid. The field redefinition is taken to be $f(r) = \sqrt{1 + a V(r)/V(0)}$ where $a$ is a constant.
The field redefinition does several things:
\begin{enumerate}
\item The potential in the center of the nucleus becomes $ V(0)(1+a) - a E $. Thus the depth of the potential is changed and becomes energy dependent. Energy dependent potentials are sometimes used. For example, deep hole states are sometimes described with a potential with a depth of about 60 MeV while 40 MeV is more common for states near the Fermi surface. The potential is not an observable.
\item It introduces an effective mass $\mathcal M(r)$. Effective masses are common in nuclear physics with values the order of $0.8m$ in the center of nuclei being common but with large variations. In the current example the central effective mass would be $m/(1+ a)$. The effective mass is not an observable.
\item The shape of the potential in the surface is changed. This is both from the $|f(r)|^2$ factor multiplying the the original potential and from derivatives acting on $f(r)$ in the original kinetic energy term. Again, the potential is not an observable.
\item The transition operator becomes $\exp[ip\cdot r]  \left[ 1 + a V(r)/V(0) \right]$ and the DWBA is no longer strictly valid. The size of the correction depends on the numerical evaluation of $a\exp[ip\cdot r] V(r)/V(0) $ and may or may not be small. As shown in the previous section, corrections to the DWBA arise naturally when an effective one-body formalism is derived from the many-body formalism. The transition operator is not an observable.
\end{enumerate}
None of these changes violates any physical principles or even common practice, although common practice would put constraints on $a$. However, these constraints are cosmetic since all the observables are independent of $a$ even when the change looks odd (the only constraint is $a>-1$ to avoid $f(r)$ being zero at some radii).  

The new equations, the equivalent of eqs.~\ref{eq:sigma} through \ref{eq:S}, can now be written:
\begin{eqnarray}
&&\sigma \propto \mathcal S\left| \int d^3r \hat\varphi_f^*(r)\mathcal T \varphi_i(r)\right |^2 \label{eq:sigmav}\\
&&-\hbar^2 \nabla\frac{1}{2 \mathcal M(r) } \nabla\varphi(r) +  \mathcal V(E,r) ] \varphi(r) = E\varphi\label{eq:schv}\\
&&\mathcal T = e^{iq\cdot r} \left(1 - \frac{d\mathcal V(E,r)}{dE}\right)\label{eq:Tv}\\
&&\mathcal S = \frac{1}{\int d^3r \hat\varphi^*(r) \left(1 - \frac{d\mathcal V(E,r)}{dE}\right) \hat\varphi(r) }\label{eq:Sv}
\end{eqnarray}
Thus we have two sets of equations, one with $V$, $\phi(r)$, $T$ and $S$ and the other with $\mathcal V$, $\varphi(r)$, $\mathcal T$ and $\mathcal S$. They both give the same results for all observables, in particular for $\sigma$, and both satisfy the consistency conditions. Phenomenologically they are equivalent. However, in general $V\neq \mathcal V$, $\phi(r) \neq \varphi(r)$, $T\neq\mathcal T$ and $S\neq \mathcal S$. Of special importance, the spectroscopic factors, $S$ and $\mathcal S$, are different. For a given state, and a given measured cross-section, the extracted spectroscopic factor can be varied from infinitesimal to infinite by the choice of $f(r)$. The actual value depends on the energy dependence we have induced into the potential by the field redefinition. The DWBA approximation is only valid for a limited range of $f$'s. We have gone beyond the pure DWBA, but only as far as it is required by gauge invariance.  As with the many-body unitary transformation, we are shuffling contributions between the wave functions and the operators.

This result is disturbing: It is very unlikely that a meaningful spectroscopic factor can be extracted from measurements. The choice of $f(r)$ gives too much freedom and there may even be other ways of generating phase equivalent potentials. However, not all choices of $f(r)$ correspond to valid many-body models. In the absence of recoil the spectroscopic factor cannot exceed one, here it can go to infinity. There is no guarantee, that even if the spectroscopic factor is less than one, that the potentials and wave functions correspond to valid many-body models. The only sure proof seems to be to calculate the potentials from a microscopic approach. This defeats the purpose of using spectroscopic factors which is at least in part to avoid the need for a microscopic calculation. As shown in the previous section, even if we do a full microscopic calculation, the spectroscopic factor still depends on the representation. 

Since $f(r)$, or the unitary transformation in the many-body calculation, can depend on the mass number using a reference nucleus does not help. The $A$ dependence can be either explicit or implicit as in the example above where $V(r)$ depends implicitly on $A$. Hence, the ratio of the spectroscopic factor in one nucleus to that in another nucleus can also be varied arbitrarily.  In the next section we will see that relative strengths in a given nucleus can also be varied if we take the field redefinition to be nonlocal. 

The dependence of the extracted spectroscopic factors on the choice of potentials is nothing new. What is new in this paper is the claim that the uncertainty due to the choice of potentials can not be eliminated by a better determination of the potentials, but is inherent in the definition of the spectroscopic factor. For example, in refs.~\cite{KBL} and \cite{IW}, a comparison is made between spectroscopic factors for the same state extracted from two different reactions. When the spectroscopic factors, as published are used, there is up to a factor of two difference. However, when the experimental data are reanalysised with a consistent set of potentials the spectroscopic factors from different reactions agree. It thus appears that spectroscopic factors can be used to transfer information from one reaction to another, but only when consistent potentials are used. Thus spectroscopic factors should always be quoted with the potentials used to extract them. 

\section{The Effective One-body Problem: The Particle Only Approach}
\label{sec:po}

If we wish to restrict the problem to particle states only, the relevant projection operator is:
\begin{eqnarray}
&&P=\int drdr'\; a^\dag(r)|\Psi_A\rangle{{\mathcal N}(r,r')}^{-1}  \langle\Psi_A |a(r')
\end{eqnarray}
where:
\begin{eqnarray}
&&{\mathcal N}(r,r')= \langle \Psi_A | a(r)a^\dag(r') | \psi_A \rangle.
 \end{eqnarray}
The projection operator here is more complicated than in the particle-hole formalism due to the norm operator, ${\mathcal N}(r,r')$. This operator plays an important role in cluster model calculations (see for example ref.~\cite{VL}), where it helps ensure that the Pauli principle is not violated. The equation for $\phi(r)$ is now:
\begin{eqnarray}
E\phi(r) &= & \int dr'dr''\; \langle\Psi|a(r)\vspace{0.5cm} \left( H + H Q \frac{1}{E - Q H Q} Q H \right) 
a^\dag(r') |\Psi\rangle{{\mathcal N}(r',r'')}^{-1}\phi(r'').\label{eq:sp}
\end{eqnarray}
Note that this equation is explicitly non-Hermitian due to the presence of the norm operator. As pointed out in ref.~\cite{VL}, this disqualifies the potential in this equation from being identified with the phenomenological optical potential. However, we can get a symmetric form by introducing the auxiliary function $\bar\phi(r)$ defined by the non-local field redefinition:
$\phi(r) =\int dr'\;{\mathcal N}(r,r')^{1/2}\bar\phi(r')$. The equation for $\bar\phi(r)$ is:
\begin{eqnarray}
E \bar\phi(r)& =& \int dr'dr''dr'''\; {{\mathcal N}(r,r')}^{-1/2} \langle\Psi|a(r') \left( H + H Q \frac{1}{E - Q  H Q} Q H \right) a^\dag(r'') |\Psi\rangle\nonumber\\&&\hspace*{9.5cm}\times{{\mathcal N}(r'',r''')}^{-1/2}
\bar\phi(r'''). \label{eq:bar}
\end{eqnarray}
This equation is explicitly symmetric but may not be Hermitian depending on how the singularity is handled. The potential in both of the preceding equations is energy dependent. The energy dependence is a hallmark of the Fechbach projection operator formalism.

\begin{figure}[t]
\includegraphics[width=13cm]{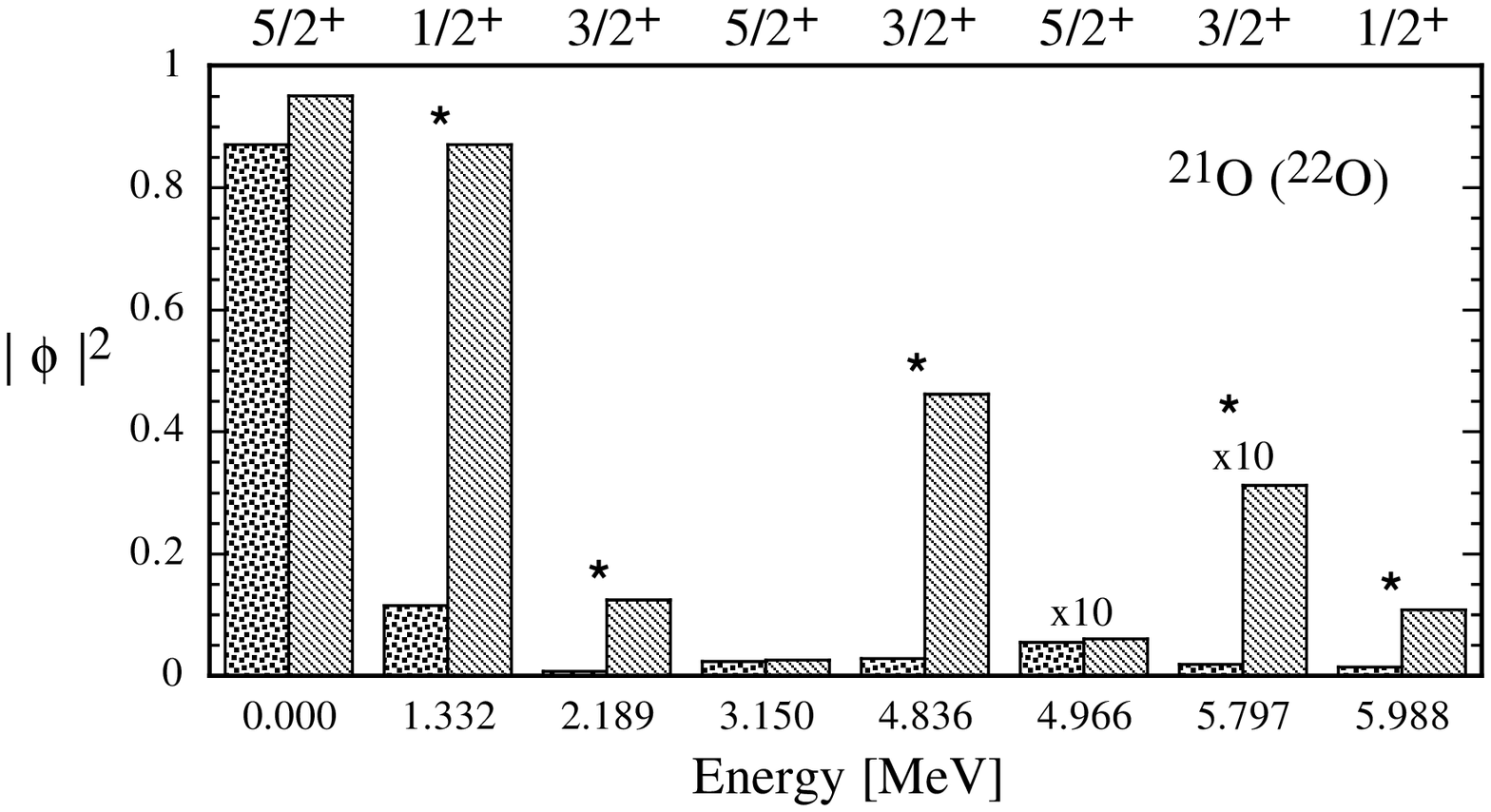}
\vspace{1cm}
\includegraphics[width=13cm]{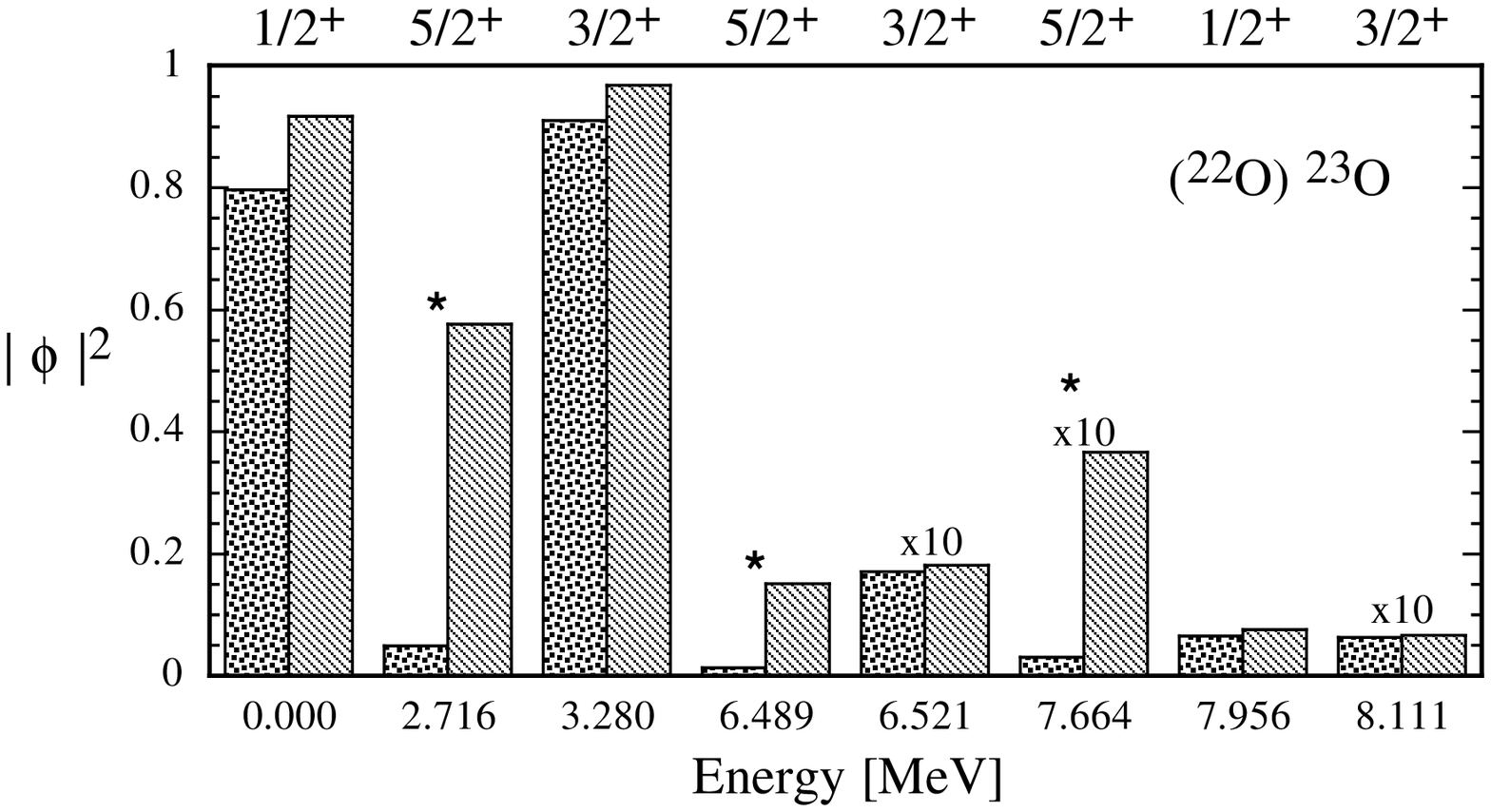}
\caption{Comparison of the spectroscopic factors and normalization of the
barred auxillary functions (dotted and striped bars respectively), for low-lying states in $^{21}$O and
$^{23}$O. All values are calculated with OXBASH using the WBP interaction
in a 2$\hbar\omega$ space~\protect{\cite{WBPint}}. 
The states marked with an asterisk are not present
in a simple one-particle model. The figures are from ref.~\cite{escher}}
\label{fig:two}
\end{figure}

It is the Hamiltonian in this last  equation that ref.~\cite{VL} identifies with 
the phenomenological Hamiltonian. This is in contrast to the identification in Sec. 2 of the present work of the particle-hole self-energy with the phenomenological Hamiltonian. The difference is significant as the normalization of $\phi(r)$ and $\bar\phi(r)$ can be quite different. For example, we show the norms for the two functions for $^{22}$O plus or minus one neutron in Fig.~\ref{fig:two}. Additional examples are given in ref.~\cite{EJ}. Notice particularly the $1/2^+$ state in $^{21}$O. In a pure single particle model it would be unoccupied and there would be no hole state. The calculated spectroscopic factor is correspondingly small. In contrast, the normalization of $\bar\phi(r)$ is almost one. Similarly, but not quite as extreme, considerations apply to the $5/2^+$ state in O$^{23}$. If the potentials  corresponding to eq.~\ref{eq:massab} are used, the norm of $\phi(r)$, the spectroscopic factor, is extract from the measured cross-sections; if eq.~\ref{eq:bar} is used the norm of $\bar\phi(r)$ is extracted. Unless we are clear which equations the phenomenological potential corresponds to the interpretation of the results could be very misleading. It is also worth noting that this non-local redefinition can change the relative normalizations in a given nucleus. In the end the conclusion is the same: extracted spectroscopic factors or normalizations are only meaningful in the context in which they are extracted. The normalizations of $\phi(r)$ and $\bar\phi(r)$ are equally unmeasurable.

\section{Conclusions}
\label{sec:con}

It has been demonstrated that spectroscopic factors are ill-defined quantities. At the level of microscopic calculations,  they can be varied by unitary transformations and we expect hard and soft nucleon-nucleon potentials to generate quite different values for them. At the phenomenological level, the spectroscopic factor can be varied by field redefinitions. Thus values extracted from experiment depend as much on the field definition as on the measured crosssection. Comparing different nuclei does not change the situation. If nonlocal field redefinitions are used, even the relative strengths in a given nucleus can change. It might be hoped that the range of variation could be constrained or conditions could be imposed to make the extracted spectroscopic factors meaningful\cite{FS}. Considering, that even if trivial phases are neglected, it takes $(N-1)^2$ real parameters to specify the unitary operators in an $N$-dimensional space, it will be difficult to sufficiently limit the variation. The author has no idea how to impose sufficient constraints in this multidimensional space. However, the combination of spectroscopic factors, optical potentials and transition operators can be used as a package to transfer information from one reaction to another.

More generally it is the asymptotic properties that tend to be observables. The phase-shift is an observable, not the wave function in the interior; the asymptotic s- to d-state ratio in the deuteron, not the d-state probability; and the asymptotic normalization coefficient\cite{mukc}, not the spectroscopic factor. Unfortunately, processes that are not strongly peaked are not uniquely determined by the asymptotic properties. Thus  the bulk quantities we frequently want are not observables while the asymptotic properties are observables but do not contain the information necessary to fully specify the reaction of interest. 

The ambiguity in the spectroscopic factor arises because contributions to the reaction crosssections can be shifted between nuclear structure and the reaction mechanism by unitary transformations or field redefinitions. The difference between the DWBA calculation and experiment can be attributed to nuclear structure, i.e. the spectroscopic factor, or the transition operator, i.e. a failure of the DWBA. This is an example of the Duhem-Quine thesis\cite{DQ}, namely that it is impossible to test a scientific hypothesis in isolation, because an empirical test of the hypothesis requires one or more background assumptions. 

There is significant collateral from spectroscopic factors not being observables. The hope was that the spectroscopic would allow one to quantify the extent to which a given state was single-particle in nature and not collective. Clearly that is not the case, but the problem goes deeper. The problem is not that spectroscopic factors are unmeasurable but that the question they are supposed to answer is poorly posed. The very concept of a single-particle state is ill defined. Just as a unitary transformation can change the spectroscopic factor, it can change the extent to which a given state is single particle or collective. It seems the best definition we can have for a single-particle state is that it is a state described by a simple single-particle model. One might object to the word ``simple", but eqs.~\ref{eq:massab}, \ref{eq:sp} and \ref{eq:bar} are all single-particle models that describe all states, certainly not simple but equally certainly single particle. It is the condition of simplicity that the selects out some states as single-particle. It is doubtful that ``simple" can be quantified sufficiently to allow an experimental determination. 
  
\section{Acknowledgements}
Financial support from the Natural Sciences and Engineering Research Council of Canada (NSERC) is appreciated. TRIUMF receives federal funding via a contribution agreement through the National Research Council of Canada. R.J.~Furnstahl is thanked for carefully reading the paper and for many useful comments.

\end{document}